\documentclass[titlepage,12pt]{utarticle}
\usepackage{amsmath,amsfonts,bbm,mathrsfs}
\usepackage{epsfig}

\newcommand{\beq}{\begin{equation}}
\newcommand{\eeq}{\end{equation}}

\begin{document}
\title{\vskip-1cm Towards Mirror Symmetry with\\ Semi-Realistic Intersecting Brane Configurations}
\author{Harun Omer}
\oneaddress{
Jefferson Physics Laboratory\\
Harvard University\\
Cambridge MA 02138, USA\\
{\tt homer@fas.harvard.edu}\\{~}\\}
\Abstract{A $T^6$ orbifold compactification is discussed from the somewhat unconventional perspective as the large radius limit of a Landau-Ginzburg model. The features of the model are in principle familiar, but the way they enter here is different from the way they enter when using more commonly used methods. It is hoped that the point of view presented here can supplement the understanding of the features used in string compactifications, notably in terms of naturalness and completeness. More precisely, the analyzed $T^6/ \mathbbm{Z}_4 \times \mathbbm{Z}_4 $ features two different kinds
of O-planes, branes in the bulk as well as fractional branes, continuous and discrete Wilson lines as well as an orientifold action which can act in different ways on the Wilson lines. The D-branes are desribed by matrix factorizations. This work is also intended to be a showcase for the potentials of matrix factorizations which are for the first time geared to their full level of sophistication in this paper.\\
Throughout the analyis everything is mapped from the B-model side of the LG-model to the A side by mirror symmetry.
The work could be extended straightforwardly yet tediously to perform mirror symmetry on a general intersecting brane configuration and to compute Yukawa couplings.\\
The analysis presented here can also be applied to non-toroidal backgrounds with an intersecting brane configuration on it, so I hope that it will be a helpful basis for later applications of mirror symmetry to models exhibiting real world properties.}
\maketitle
\section{Introduction}
Intersecting D-branes on toroidal backgrounds have been the object of intensive study for several years now and have been used successfully in the construction of semi-realistic models which are able to reproduce the standard model gauge group, the chiral fermionic particle spectrum and other aspects of the physical world. Two good reviews are~\cite{blumenhagenreview,marchesanoreview}. T-duality relates the type IIA and type IIB string theories in toroidal models.\\
There are some limitations to the conventional methods used, notably the difficulties when dealing with non-factorizable cycles.
But first and foremost is the restriction to only toroidal models. Within the topological context, this limitation can
be lifted. The generalization of T-duality to a correspondence between general Calabi-Yau manifolds is achieved by homological mirror symmetry. Mirror symmetry is a statement of equivalence between a Landau-Ginzburg model with a Calabi-Yau manifold $\mathcal{X}$ as target space and a sigma-model with some other Calabi-Yau $\mathcal{Y}$ as target space.
By introducing a boundary on the world-sheet it is possible to include D-branes. Matrix factorizations~\cite{matfac1,matfac2} have developed to a practicable tool to describe the various possible boundary conditions corresponding to different D-branes. (See e.g.~\cite{jockerslerche} and references therein to get acquainted with the subject.)\\
It has never been attempted to perform mirror symmetry on a fully-fledged semi-realistic string compactification; perhaps it was thought to be beyond practical feasibility. The goal of this work is to show from an example how it can be done, although I do not pursue this until the end and stop at the point where it is clear how to proceed. The example used is the $T^6 / \mathbbm{Z}_4 \times \mathbbm{Z}_4$ orientifold. Since mirror symmetry is a generalization of T-duality, namely a T-duality on a torus fibration of a Calabi-Yau, it might seem we are back at where we started. Ultimately this must of course be true, but the methods used are more general. T-duality is never refered to in this paper; a Landau-Ginzburg superpotential which 'happens' to describe the toroidal orbifold is the starting point, but specific features of the potential are not used in the calculation. It should be possible to perform an analogous calculation with a potential describing a proper Calabi-Yau manifold in which other researchers might be more interested in. The advantage of the toroidal model is that results are already known and as such it provides a good test case. In addition, the computation gives an entirely new point of view on toroidal compactifications. After all, Landau-Ginzburg models are defined in the small radius limit, thus probing stringy geometry. Only in the large radius limit does the model discussed here obtain an interpretation as a toroidal compactification. As such, even from the toroidal case things can be learnt. Features appear from {\it within} the theory naturally. It will be shown, how up to the last subtlety every feature of orientifold compactifications are show up. In particular, two different kinds of O-planes can be chosen from; bulk cycles are present and so are fractional branes; continous as well as discrete Wilson lines emerge naturally; and the orientifold action can act in different consistent ways on the Wilson lines. In addition it can be seen how the LG-target space of this topological model can in some sense be regarded as a deformed orbifold.\\
In principle it is possible without further fundamental difficulties to reproduce for example semi-realistic $T^6/\mathbbm{Z}_2 \times \mathbbm{Z}_2$ orientifold spectra considered in the literature, compute their Yukawa couplings and the low-energy effective action and perform mirror symmetry on the entire configuration. However, there is no point in reproducing known results. The purpose of the paper is merely to show how certain computations can be performed.  
Work to derive new results for the toroidal orbifold is currently underway.\\
Note also that all kinds of different cristallographic orbifold actions have been discussed in the literature. An exception is the group $\Gamma=\mathbbm{Z}_4 \times \mathbbm{Z}_4$ which according to~\cite{forste} is perturbatively inconsistent. The methods used here are non-perturbative, which allows to deal with this case. It should also be remembered, that all geometric notions are only valid interpretations in the large radius limit. The Landau-Ginzburg model is a priory a small-radius (strong coupling) theory where geometry is 'blurred' by quantum effects. Nevertheless, I sloppily refer to cycles, dimensions and toroidal orbifolds all the time in reference to the interpretation of the large radius limit.\\
Furthermore note that all steps can be repeated easily for other Calabi-Yau manifolds except for one important open question: There is no systematic way yet to find the flat coordinate system which is needed to pursue an analyis like the one presented below.    
\section{The $T^6$ orientifold}
The starting point is the Landau-Ginzburg model with superpotential,
\begin{eqnarray}
W=\sum_{i=1}^3 x_{2i-1}^4+x_{2i}^4-a_i x_{2i-1}^2 x_{2i}^2 - z_1^2 \;\;(-z_2^2 -z_3^2).
\end{eqnarray}
For convenience the irrelevant squared terms in brackets are sometimes added. This issue is addressed further below.
In the large radius limit, the theory is interpreted as type IIB string theory on a (orbifolded) $T^6$ target space with the deformations $a_1(\tau_1)$, $a_2(\tau_2)$ and $a_3(\tau_3)$ parametrizing the three complex structures of the tori.  All other deformations correspond to blow-up moduli and are assumed to be small. Mirror symmetry maps the theory to the type IIA side, thereby exchanging the complex structure and K\"ahler moduli. On the IIA side, the theory corresponds to a torus orbifold $T^6 / \mathbbm{Z}_4 \times \mathbbm{Z}_4$.
The $T^6$ we are dealing with is factorizable, which means that a decomposition into two-tori is respected by the imposed orbifold and orientifold actions. 
In the following, first the two-dimensional case is reviewed.
\section{The moduli-dependent $T^2$ torus}
\subsection{Three-variable case}
In~\cite{knappomer2} homological mirror symmetry on the $T^2$ described by the quartic curve in projective space was discussed. The LG superpotential,
\begin{eqnarray}
W^{(3)}=x_1^4+x_2^4-a x_1^2 x_2^2 - z_1^2,
\label{eq:T2pot}
\end{eqnarray}
includes the extra square term $z_1^2$ so that the fermion number is equal to the central charge $\hat{c}$ mod 2. The deformation parameter $a=a(\tau)$ is related to the torus complex structure modulus $\tau$ in terms of the torus modular invariant function~\cite{Giveon:1990ay},
\begin{equation}
\label{jrel}
j(\tau)=\frac{16(a^2+12)^3}{(a^2-4)^2}.
\end{equation}
The model, which can in principle be solved directly on the CFT level, corresponds to the point $\rho=e^{2 \pi i/4}$ in K\"ahler moduli space, defining a square torus lattice $\mathbbm{C}/(\mathbbm{Z} \times \rho \mathbbm{Z})$ on the mirror $A$-side. In order to be able to actually perform mirror symmetry, flat coordinates have to be used, which means that the moduli have to be parametrized in a particularly natural manner. Sections of line bundles $\alpha_{1,2,3}\equiv \alpha_{1,2,3}(u,\tau)$ depending on some boundary modulus $u$ are introduced. They must themselves satisfy the surface equation $W=0$:
\begin{eqnarray}
\alpha_1^4+\alpha_2^4-a\; \alpha_1^2 \alpha_2^2 - \alpha_3^2=0. \label{eq:thetarel}
\end{eqnarray}
In terms of these sections, a minimal basis of matrix factorizations $Q^{(3)}_k$ was found in~\cite{knappomer2}.
Explicitly, the branes are represented by,
\begin{eqnarray}
Q^{(3)}_k=\left(\begin{array}{cc}0&E_k\\J_k&0\end{array}\right),\label{eq:3x3fac}
\end{eqnarray}
\begin{equation}
\label{short1}
E_k=\left(\begin{array}{cc}
\alpha_1^ix_1+\alpha_2^ix_2& z_1+\frac{\alpha_3^k}{\alpha_1^k\alpha_2^k}x_1x_2\\
-z_1+\frac{\alpha_3^k}{\alpha_1^k\alpha_2^k}x_1x_2&-\frac{1}{\alpha_1^k}x_1^3+\frac{\alpha_1^k}{(\alpha_2^k)^2}x_1x_2^2-\frac{1}{\alpha_2^k}x_2^3+\frac{\alpha_2^k}{(\alpha_1^k)^2}x_1^2x_2
\end{array}\right),
\end{equation} 
\begin{equation}
\label{short2}
J_k=\left(\begin{array}{cc}
\frac{1}{\alpha_1^k}x_1^3-\frac{\alpha_1^k}{(\alpha_2^k)^2}x_1x_2^2+\frac{1}{\alpha_2^k}x_2^3-\frac{\alpha_2^k}{(\alpha_1^k)^2}x_1^2x_2&z_1+\frac{\alpha_3^k}{\alpha_1^k\alpha_2^k}x_1x_2\\
-z_1+\frac{\alpha_3^k}{\alpha_1^k\alpha_2^k}x_1x_2&-\alpha_1^k x_1-\alpha_2^k x_2
\end{array}\right).
\end{equation}
The label $k$ for the four orbifold copies is suppressed below for better overview. 
On the B-side the four orbifold copies $k=0,...,3$ of the factorization correspond to a pure D0 and a pure D2 brane as well as their anti-branes (which in this case are isomorphic to the branes themselves). This can be shown by computing the ranks and degrees of the bundle, which are $(r,c_1)=(0,1)$ and $(r,c_1)=(1,0)$ for the two orbifold branes. On the (unorbifolded) mirror A-side these numbers become the wrapping numbers of branes wrapping the two fundamental 1-cycles of the $T^2$.
Their location on the torus can be read off from the respective boundary modulus $u$ provided that the parameterization $\alpha_{1,2,3}$ is in the flat coordinate basis suitable for mirror symmetry. It was argued that the appropriate sections are certain theta functions,
\begin{eqnarray}
\label{uniformization}
\begin{array}{l}
\displaystyle \alpha_1(u,\tau)=\Theta_2(2u,2\tau)\qquad \alpha_2(u,\tau)=\Theta_3(2u,2\tau),\\
\displaystyle \alpha_3(u,\tau)=\frac{\Theta_4^2(2\tau)}{\Theta_2(2\tau)\Theta_3(2\tau)}\Theta_1(2u,2\tau)\Theta_4(2u,2\tau),
\end{array}
\end{eqnarray}
where $\Theta_i(\tau)\equiv\Theta_i(0,\tau)$. Eq.~(\ref{uniformization}) differs from the one given in~\cite{knappomer2} by a shift of the origin.\\
The deformation parameter $a(\tau)$ which is subject to relation~(\ref{jrel}) can also be expressed in terms of theta functions,
\begin{equation}
a(\tau)=\frac{\Theta_2^4(2\tau)+\Theta_3^4(2\tau)}{\Theta_2^2(2\tau)\Theta_3^2(2\tau)}.
\end{equation}
These branes $Q^{(3)}_k$ can be moved around continously by virtue of their boundary modulus $u$.
\subsection{Two-variable case}
Apart from the model just described, the LG potential Eq.~(\ref{eq:T2pot}) without the additional squared term $z_1^2$ was also discussed in~\cite{knappomer2}. While the geometrical interpretation of this model is less clear, it has the advantage
that the potential can be rewritten in a simple product form,
\begin{equation}
W^{(2)}=x_1^4+x_2^4 - a x_1^2 x_2^2 = \overset{4}{\underset{n=1}{\displaystyle \prod}} (x_1-\eta_n x_2). 
\end{equation}
The four coefficients are,
\begin{equation}
\label{etadef}
\eta_{n}=\pm \sqrt{\frac{a}{2} \pm\sqrt{{\left(\frac{a}{2}\right)^2-1}}}\qquad n\in D=\{1,2,3,4\},
\end{equation}
and become fourth roots of unity at the Gepner point. I use the convention $\eta_1\simeq (+,+),
\eta_2\simeq (+,-),\eta_3\simeq (-,+),\eta_4\simeq (-,-)$.
In this two-variable model we have some simple $1\times 1$ factorizations,
\begin{eqnarray}
Q^A=\begin{pmatrix}0&E^A\\J^A&0\end{pmatrix}\qquad
E^A=\underset{n \in I_A}{\displaystyle \prod} (x_1-\eta_n x_2)\qquad
J^A=\underset{n \in D \backslash I_A}{\displaystyle \prod} (x_1-\eta_n x_2).
\end{eqnarray}
with a spectrum which is derived easily. The index set $I_A$ is a subset of $D$ and the norm of it is defined to stand for the number of elements it contains. The four coefficients $\eta_n$ are related through,
\begin{eqnarray}
\eta_{1,3}=\pm \frac{\alpha_1}{\alpha_2}\qquad
\eta_{2,4}=\pm \frac{\alpha_2}{\alpha_1},\label{eq:coeff}
\end{eqnarray}
to the parametrization of the surface equation,
\begin{eqnarray}
\alpha_1^4+\alpha_2^4-a\; \alpha_1^2 \alpha_2^2=0.\label{eq:2varjac}
\end{eqnarray}
For this model, however, no sections $\alpha_{1,2}(u,\tau)$ compatible with Eq.~(\ref{eq:2varjac}) exist: The $\eta_n$ were derived by regarding $W=0$ as a fourth order polynomial equation in $x_1$, so from the fundamental theorem of algebra it is clear that there are no further zeros, let alone a continous zero locus.\\
The independence from the modulus can also be seen by recombining two of the permutation type branes of the form,
\begin{eqnarray}
Q^{(2)}=\begin{pmatrix} 0 & E\\ J & 0\end{pmatrix},\qquad
E=(x_1-\eta_n x_2)\qquad
J=\underset{m \ne n}{\displaystyle \prod} (x_1-\eta_m x_2).
\label{eq:2x2perm}
\end{eqnarray}
By tachyon condensation, one obtains with the help of Eq.~(\ref{eq:2varjac}) and a similarity transformation a resulting brane whose boundary modulus dependence has dropped out explicitly,
\begin{equation}
\tilde Q=\begin{pmatrix}0&\tilde E\\ \tilde J&0\end{pmatrix},\qquad
\tilde E=\begin{pmatrix}x_2^3& x_1^3-2 a x_1 x_2^2\\ x_1&-x_2\end{pmatrix},
\tilde J=\begin{pmatrix}x_2& x_1^3-2 a x_1 x_2^2\\ x_1&-x_2^3\end{pmatrix}.
\nonumber
\end{equation}
One can also start with a formal modulus dependance and then show that the boundary operator
$\langle\Omega\rangle=\langle \partial_{u} Q \rangle$ vanishes regardless of the form of the $u$-dependance of $\alpha_{1,2}$. For this one uses the Kapustin-Li formula,
\begin{equation}
\langle\Omega\rangle=\int\frac{\mathrm{STr}\left(\frac{1}{2!}(\mathrm{d}Q)^{\wedge 2}\partial_u Q \right)}{\partial_1W\partial_2W}.
\end{equation}
Only at the discrete points where $\alpha_3(u,\tau)$ of Eq.~(\ref{uniformization}) vanishes, the Jacobian of the three-variable potential Eq.~(\ref{eq:thetarel}) becomes the Jacobian of the two-variable potential Eq.~(\ref{eq:2varjac}). These discrete points are the zeros in $u$ of $\Theta_1(u,\tau)$ and $\Theta_4(u,\tau)$. 
\section{A Minimal Set of Branes}
\subsection{Intersection Numbers}
From the Witten index we know the intersection numbers between
D-branes and can establish whether we are dealing with a minimal basis of the charge lattice. A unimodular intersection form indicates an integral basis of the free module. The index theorem reads~\cite{Walcher:2004tx},
\begin{eqnarray}
\mbox{Tr}(-1)^F=\frac{1}{H} \sum_{k=1}^{H-1}\mbox{STr}(\gamma_P^k)\frac{1}{\Pi_i(1-\omega_i^k)}\mbox{STr}(\gamma_Q^{-k}),
\label{eq:intnum}
\end{eqnarray}
where $\gamma_Q$ denotes the orbifold generator on the factorization $Q$, the product index $i$ runs over the number of fields in the LG-action, $q_i$ is their R-charge, appearing in $\omega_i=e^{i \pi q_i}$ and $H=4$ for our $\mathbbm{Z}_4$-orbifold.
The orbifold generators $\gamma_{(2)}$ and $\gamma_{(3)}$ of the factorization of the $W^{(2)}$ and $W^{(3)}$ potentials are,
\begin{eqnarray}
\gamma_{(2)}=i^{n} \begin{pmatrix}1 & 0\\ 0 & i\end{pmatrix}\;\;\;\;
\gamma_{(3)}=i^{n} \begin{pmatrix}i & 0 & 0 & 0\\0 & 1 & 0 & 0\\ 0 & 0 &-1 & 0\\ 0 & 0 & 0 & -i\end{pmatrix} \;\; n=0,1,2,3.
\end{eqnarray}
Taking the supertrace,
\begin{eqnarray}
\mbox{STr}(\gamma_{(3)}^k)=i^n \,2(1+i^k),\label{eq:D0orb}
\end{eqnarray}
we find the intersection numbers of the two orbifold copies of the $W^{(3)}$-factorization, 
\begin{eqnarray}
\mathbb{I}^{(3)}_{nm}=\begin{pmatrix} 0 & 1\\-1 & 0 \end{pmatrix},
\end{eqnarray}
confirming the intersection number we know from the A-side. In order to construct a $T^6$ orbifold, it would seem most natural to tensor together three copies of $W^{(3)}$.
The factorization $Q^{(3)}$ was a 1-cycle wrapping the $T^2$ on the topological $A$-side and if we denote the two 
fundamental cycles on the covering space by $\pi_1$ and $\pi_2$ the two orbifold copies represent these cycles. The other two copies are identified with these cycles as well and can therefore be disregarded. Note that by imposing the orbifold condition on the B-side, one actually unorbifolds on the A-side. Since the $T^2$ Landau-Ginzburg model is subject to only one $\mathbbm{Z}_4$-symmetry, we obtain an unorbifolded $T^2$ on the A-side. This changes in the tensored model. There we have a $\mathbbm{Z}_4^3$-symmetry, and imposing the orbifold condition of the $\mathbbm{Z}_4$ quantum symmetry, we are left with a
$\mathbbm{Z}_4\times \mathbbm{Z}_4$-orbifold model. Consequently, tensoring together three copies of the brane $Q^{(3)}$ must lead to a brane which wraps fundamental cycles on the three tori and is invariant under the orbifold action. Denoting the two
fundamental cycles on the n-th torus as $\pi_{2n-1}$ and $\pi_{2n}$, the type IIA 3-cycles can be written,
\begin{eqnarray}
\pi_{klm}\equiv \pi_k \otimes \pi_l \otimes \pi_m.
\end{eqnarray}
Geometrically one finds that there are two orbifold invariant combinations:
\begin{eqnarray}
\Pi_1 = 2(\pi_{135} - \pi_{236} -\pi_{146}-\pi_{245}),\\
\Pi_2 = 2(\pi_{136} - \pi_{235} -\pi_{145}-\pi_{246}).
\end{eqnarray}
They differ by a rotation of $\pi/2$ on each $T^2$. and can be identified easily with the two orbifold copies of the tensored branes.\\
Let us forget about our topological model for a moment and make contact with standard methods to make things clearer.
The generators $\theta^{1,2}$ of the orbifold group $\mathbbm{Z}_4 \times \mathbbm{Z}_4$ act on the complex coordinates $z^i=x^i+\tau^i y^i$ of $T^6$ as,
\begin{eqnarray}
\theta^{1,2}: (z^1,z^2,z^3) \longrightarrow (e^{2\pi i v_1} z^1,e^{2\pi i v_2} z^2,e^{2\pi i v_3} z^3),\label{eq:orbi}
\end{eqnarray}
where $v^1=\frac{1}{4}(1,0,-1)$ and $v^2=\frac{1}{4}(0,1,-1)$.
The cohomology class $H_3(T^6,\mathbbm{Z})$ of the torus is a priori 20-dimensional. The orbifold condition Eq.~(\ref{eq:orbi}) projects out all but the $(3,0)$ and $(0,3)$ components which survive every orbifolding: $\Omega_3=dz^1 \wedge dz^2 \wedge dz^3$ and
$\overline \Omega_3=d\overline{z}^1 \wedge d\overline{z}^2 \wedge d\overline{z}^3$. $\Pi_{1,2}$ correspond the linear combinations $\mbox{Re }2\Omega_3$ and $2\mbox{Im }\Omega_3$ of the 3-forms. Expanded they read,
\begin{eqnarray}
\begin{array}{l}
\mbox{Re }2\Omega_3=\Omega_3+\overline \Omega_3=\\
2\underbrace{dx^1\wedge dx^2\wedge dx^3}_{\pi_{135}}-2\underbrace{dx^1\wedge dy^2\wedge dy^3}_{\pi_{146}}-2\underbrace{dy^1\wedge dx^2\wedge dy^3}_{\pi_{236}}-2\underbrace{dy^1\wedge dy^2\wedge dx^3}_{\pi_{245}},\nonumber\\
\mbox{Im }2\Omega_3=-i(\Omega_3-\overline \Omega_3)=\\
2\underbrace{dx^1\wedge dx^2\wedge dy^3}_{\pi_{136}}-2\underbrace{dx^1\wedge dy^2\wedge dx^3}_{\pi_{145}}-2\underbrace{dy^1\wedge dx^2\wedge dx^3}_{\pi_{235}}-2\underbrace{dy^1\wedge dy^2\wedge dy^3}_{\pi_{246}}.
\end{array}\nonumber
\end{eqnarray}
The intersection matrix computed from the IIA side by integration over the cycles is,
\begin{eqnarray}
\frac{1}{4}\Pi_n\circ \Pi_m=\begin{pmatrix} 0 & -4\\4 & 0 \end{pmatrix}.\label{eq:IIAint}
\end{eqnarray}
From the IIB side computation the intersection matrices for two and three branes tensored together are,
\begin{eqnarray}
\mathbb{I}^{(3)(3)}_{nm}=\begin{pmatrix} -2 & 0\\0 & -2 \end{pmatrix}\qquad
\mathbb{I}^{(3)(3)(3)}_{nm}=\begin{pmatrix} 0 & -4\\4 & 0 \end{pmatrix}.
\end{eqnarray}
The latter matrix of course reproduces the IIA intersection number Eq.~(\ref{eq:IIAint}). These branes do not span a minimal basis. But by taking the tensor product $Q^{(3)} \otimes Q^{(2)} \otimes Q^{(2)}$ instead, we obtain the desired unimodular intersection form,
\begin{eqnarray}
\mathbb{I}^{(3)(2)(2)}_{nm}=\begin{pmatrix} 0 & -1\\1 & 0 \end{pmatrix}.
\end{eqnarray}
The same is true for the usual permutation-type constructions,
\begin{eqnarray}
Q^{(2)} \otimes Q^{(2)} \otimes \begin{pmatrix} 0 & -z\\z & 0 \end{pmatrix},
\end{eqnarray}
and
\begin{eqnarray}
Q^{(2)} \otimes Q^{(2)} \otimes Q^{(2)} \otimes \begin{pmatrix} 0 & -z\\z & 0 \end{pmatrix}.
\end{eqnarray}
In the IIA-picture tachyon condensation of two branes leads to a new brane whose wrapping numbers are the sum of the wrapping numbers of the condensed branes. So with the tensored factorizations $Q^{(3)} \otimes Q^{(3)} \otimes Q^{(3)}$ one can generate the orbifold invariant bulk 3-cycles. The permutation type constructions correspond to the fractional branes. In the IIA picture, a tensor brane of type $Q^{(3)} \otimes Q^{(2)} \otimes Q^{(2)}$ represents a bulk cycle on one of the tori, sitting at fixed points of the other two tori in the orbifold limit. Under a "blow-up" of the orbifold fixed points to $S^2$ -- which could be done by switching on the corresponding perturbations in the LG potential -- the branes could be regarded as wrapping so-called exceptional 3-cycles of topology $S^2 \times S^1$.
\subsection{Inequivalent Factorizations}
In the $\eta$-notation, the most general basic three-variable factorizations read,
\begin{eqnarray}
\begin{array}{l}
E^A_k=\left(\begin{array}{cc}
\underset{n \in I_A} \prod{(x_1-\eta^k_n x_2)}& z_1+\frac{\alpha_3^k}{\alpha_1^k\alpha_2^k}x_1x_2\\
-z_1+\frac{\alpha_3^k}{\alpha_1^k\alpha_2^k}x_1x_2 &
-\underset{m \in D \backslash I_A}\prod{(x_1-\eta^k_m x_2)}
\end{array}\right),\\
J^A_k=\left(\begin{array}{cc}
\underset{m \in D \backslash I_A}\prod{(x_1-\eta^k_m x_2)} & z_1+\frac{\alpha_3^k}{\alpha_1^k\alpha_2^k}x_1x_2\\
-z_1+\frac{\alpha_3^k}{\alpha_1^k\alpha_2^k}x_1x_2&-\underset{n \in D \backslash I_A}\prod{(x_1-\eta^k_n x_2)}
\end{array}\right).
\end{array}
\label{eq:2x2general}
\end{eqnarray}
For $|I_A|=1$ these are the permutation type branes discussed, which wrap the fundamental 1-cycles in the IIA picture
and for $|I_A|=2$ they are condensed branes of the two different fundamental cycles. Up to a shift in origin, they therefore wrap the diagonals of the covering space. For details see~\cite{knappomer2}.\\
The R-charge matrices associated with factorizations with $|I_A|=1$ and $|I_B|=2$ are,
\begin{eqnarray}
R^A&=&\mbox{diag}\left(\frac{1}{4},-\frac{1}{4},-\frac{1}{4},\frac{1}{4}\right),\\
R^B&=&\mbox{diag}\left(0,0,0,0\right).
\end{eqnarray}
They define the orbifold generators:
\begin{eqnarray}
\gamma_k^A&=&e^{\frac{k \pi i}{2}}\mbox{diag}\left(e^{\frac{\pi i}{2}},1,1,e^{\frac{\pi i}{2}}\right)\qquad k=0,1,2,3,\\
\gamma_k^B&=&e^{\frac{k \pi i}{2}}\mbox{diag}\left(1,1,-1,-1 \right)\qquad k=0,1,2,3.
\end{eqnarray}
It is well-known that factorizations which can be transformed into each other by a similarity transformation describe the same brane. Naively, one would expect to obtain four different branes $Q^{(3)}$ -- one for each $\eta_n$ -- since these factorizations are inequivalent with respect to similarity transformations. However, we are here dealing with four continous families of factorizations, each defined over the moduli-space $\mathbb{C}/(\mathbb{Z}+\tau \mathbb{Z})$ and they are merely different parametrizations of the moduli-space as can be seen by some theta function identities.
From the quasi periodicity of the theta-functions,
\begin{eqnarray}
\begin{array}{rcr}
\Theta_1(2u,2\tau)&=&-\Theta_1(2(u+\frac{1}{2}),2\tau),\\
\Theta_2(2u,2\tau)&=&-\Theta_2(2(u+\frac{1}{2}),2\tau),\\
\Theta_3(2u,2\tau)&=&\Theta_3(2(u+\frac{1}{2}),2\tau),\\
\Theta_4(2u,2\tau)&=&\Theta_4(2(u+\frac{1}{2}),2\tau),
\end{array}
\end{eqnarray}
together with Eq.~(\ref{uniformization}) we find,
\begin{eqnarray}
\frac{\alpha_1(u,\tau)}{\alpha_2(u,\tau)}&=&-\frac{\alpha_1(u+\frac{1}{2},\tau)}{\alpha_2(u+\frac{1}{2},\tau)},\\
\frac{\alpha_3(u,\tau)}{\alpha_1(u,\tau)\alpha_2(u,\tau)}&=&\frac{\alpha_3(u+\frac{1}{2},\tau)}{\alpha_1(u+\frac{1}{2},\tau)\alpha_2(u+\frac{1}{2},\tau)}.
\end{eqnarray}
This corresponds to the exchange $\eta_1\longleftrightarrow \eta_3$ and $\eta_2\longleftrightarrow \eta_4$ in the brane of Eq.~(\ref{eq:2x2general}). Another internal symmetry of Eq.~(\ref{uniformization}) is a sign change in $\alpha_3$. One can show that it can be undone on the level of the branes by a reflection of the boundary modulus,
\begin{eqnarray}
\frac{\alpha_1(u,\tau)}{\alpha_2(u,\tau)}&=&\frac{\alpha_1(-u,\tau)}{\alpha_2(-u,\tau)},\\
\frac{\alpha_3(u,\tau)}{\alpha_1(u,\tau)\alpha_2(u,\tau)}&=&-\frac{\alpha_3(-u,\tau)}{\alpha_1(-u,\tau)\alpha_2(-u,\tau)}.
\end{eqnarray}
A further identity, which together with the previous one swaps $\eta_1\longleftrightarrow \eta_2$ and $\eta_3\longleftrightarrow \eta_4$ reads,
\begin{eqnarray}
\frac{\alpha_1(u,\tau)}{\alpha_2(u,\tau)}&=&\frac{\alpha_2(u+\frac{1}{2}+\frac{\tau}{2},\tau)}{\alpha_1(u+\frac{1}{2}+\frac{\tau}{2},\tau)},\\
\frac{\alpha_3(u,\tau)}{\alpha_1(u,\tau)\alpha_2(u,\tau)}&=&-\frac{\alpha_3(u+\frac{1}{2}+\frac{\tau}{2},\tau)}{\alpha_1(u+\frac{1}{2}+\frac{\tau}{2},\tau)\alpha_2(u+\frac{1}{2}+\frac{\tau}{2},\tau)}.
\end{eqnarray}
Therefore all these apparently different representations just parametrize the $T^2$ moduli space differently and it is sufficient to pick only the single factorization Eq.~(\ref{eq:3x3fac}). Alternatively it would be possible to restrict the moduli space from the entire torus covering space to the sector $\lambda_1+\lambda_2 \tau \in \mathbbm{C}$, $0 \le \lambda_i \le \frac{1}{2}$ and stick with all four different possibilites of Eq.~(\ref{eq:coeff}). They would then parametrize the untwisted and the three twisted sectors of the orbifold covering space. 
\subsection{Identifying the Fractional Branes}
\subsubsection{Type IIB side}
Mirror symmetry is a duality between an orbifolded Landau-Ginzburg model and a sigma model which is not orbifolded or vice versa. In the large radius limit these models should correspond to the IIA and IIB models. I had found it puzzling that the mirror pairs of IIA and IIB toroidal compactifications are both orbifolds (or both not orbifolds) whereas the LG-/sigma model correspondence is a duality of theories with an orbifolded and an unorbifolded target space respectively. The purpose of this section is to shed some light on this matter.\\
I start by addressing the question where the fractional branes are located.
It is easy to see from Eq.~(\ref{eq:intnum}) that intersection numbers never change when two quadratic terms $W^{trivial}=-z_1^2-z_2^2$ are added to the LG-potential and the factorizations are tensored with,
\begin{eqnarray}
Q^{trivial}=\begin{pmatrix}0 & -(z_1+i z_2)\\(z_1-i z_2) & 0\end{pmatrix}.
\end{eqnarray}
This trivial factorization $Q^{trivial}$ is completely rigid and can not be deformed. It would be possible to perform all computations without this additional term but at times it is preferable to add it for symmetry reasons since it nicely exhibits the tensor product structure of the three $T^2$. By tachyon condensation one obtains a factorization,
\begin{eqnarray}
\begin{pmatrix}0 & z_1\\-z_1 & 0\end{pmatrix} \otimes \begin{pmatrix}0 & z_2\\-z_2 & 0\end{pmatrix}.
\end{eqnarray}
With this part tensored to the fractional brane, the resulting factorization can be deformed continously (in the sense that the brane corresponds to a particular point in a continous boundary moduli space). In other words, this is the branch cut of the Coulomb branch of the fractional brane where its moduli space degenerates to a point. In order to find the locations of this brane on the target space, all one needs to do is find a bulk cycle of type $Q^{(3)} \otimes Q^{(3)} \otimes Q^{(3)}$ which for some moduli becomes identical to our condensated brane of type,
\begin{eqnarray}
Q^{(3)} \otimes Q^{(2)} \otimes  \begin{pmatrix}0 & z_2\\-z_2 & 0\end{pmatrix} \otimes Q^{(2)} \otimes
\begin{pmatrix}0 & z_3\\-z_3 & 0\end{pmatrix}.
\end{eqnarray}
The resulting fixed values of the moduli should then encode the branes location.
Obviously, one factor $Q^{(3)}$ can be identified immediately. And,
\begin{eqnarray}
Q^{(2)} \otimes  \begin{pmatrix}0 & z_2\\-z_2 & 0\end{pmatrix},
\end{eqnarray}
also has the structure of $Q^{(3)}$ at the special point where $\alpha_3$ vanishes (see Eq.~(\ref{short1}-\ref{short2})).
As said before, these are the zeros in $u$ of $\Theta_1(2u,2\tau)$ or $\Theta_4(2u,2\tau)$. From the product representation of the theta matrices, the zeros are easy to derive. There are precisely four solutions so there is no ambiguity in the assignment of the moduli to the constants $\eta_n$. In the conventions used here, the factor,
\begin{eqnarray}
E=\left(x_1 +\frac{\alpha_2(u,\tau)}{\alpha_1(u,\tau)} x_2\right),
\end{eqnarray}
must correspond to $(x_1-\eta_n x_2$). The identification goes as follows:
\begin{eqnarray}
\begin{array}{cc}
\displaystyle\frac{\alpha_2(0,\tau)}{\alpha_1(0,\tau)}=-\eta_4, &
\displaystyle\frac{\alpha_2(\frac{1}{2},\tau)}{\alpha_1(\frac{1}{2},\tau)}=-\eta_2,\\
\\
\displaystyle\frac{\alpha_2(\frac{1}{2}\tau,\tau)}{\alpha_1(\frac{1}{2}\tau,\tau)}=-\eta_1, &
\displaystyle\frac{\alpha_2(\frac{1}{2}+\frac{1}{2}\tau,\tau)}{\alpha_1(\frac{1}{2}+\frac{1}{2}\tau,\tau)}=-\eta_3.
\end{array}\label{eq:fracident}
\end{eqnarray}
\subsubsection{Quantum Orbifold Action}
\begin{figure}
\begin{center}
\includegraphics{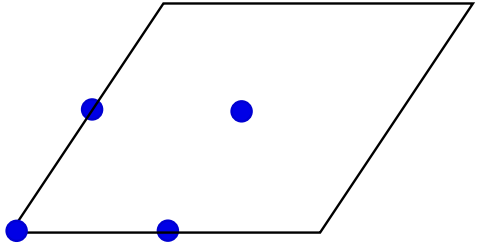}
\end{center}
\caption{The IIB covering space with the locations of the four rigid D0-brane components.}\label{fig:tilted}
\end{figure}
In Fig.~\ref{fig:tilted}, the IIB covering space of one torus is shown with the rigid branes marked on it. Usually one constructs orbifolds by imposing the orbifold condition on the flat space, fixing thereby the complex structure of the torus. A $\mathbbm{Z}_4$-orbifolded torus can therefore only have a rigid square covering space. But for $\tau=i$, Fig.~\ref{fig:tilted} places the immovable D0-branes precisely at the two $\mathbbm{Z}_4$-fixed points and the two $\mathbbm{Z}_2$-fixed points which are exchanged by the $\mathbbm{Z}_4$-symmetry. For a generic complex structure $\tau$ one can therefore regard this construction as a generalization of the conventional orbifold constructions. The underlying reason is that the $\mathbbm{Z}_4$-symmetry acts on the quasi-homogeneous LG potential $W$ without interfering with the complex structure.
\subsubsection{Type IIA side}
\begin{figure}
\begin{center}
\includegraphics{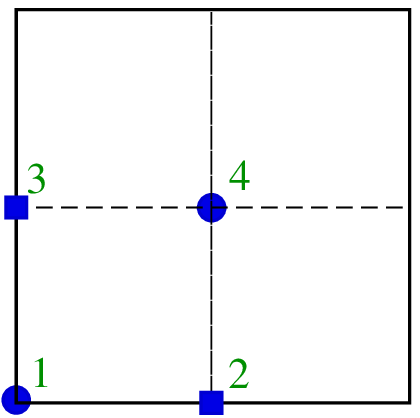}
\includegraphics{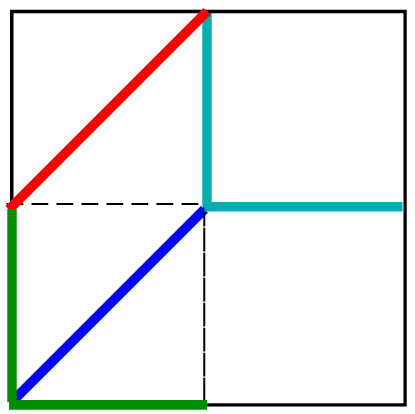}
\end{center}
\caption{The IIA covering space with the locations of the fixed points. Fractional branes drawn on the untwisted sector.}\label{fig:fixedpoints}
\end{figure}
Let us return to the fractional branes and identify them on the IIA side. Fractional branes wrap homology cycles which are a sum of half a bulk cycle and an exceptional cycle. When the orbifold singularities are "blown-up", they topologically become $S^2$ which can be wrapped by an exceptional 2-cycle. These can be tensored with a fundamental cycle on one torus.
The orbifold singularities are labeled on each torus as shown in Fig.~\ref{fig:fixedpoints}. Points 1 and 4 are $\mathbbm{Z}_4$ fixed points, points 2 and 3 are only fixed under a $\mathbbm{Z}_2$ symmetry and are exchanged under $\mathbbm{Z}_4$. A fractional brane that goes through one of these two points must necessarily also pass through the other to be invariant.
In Fig.~\ref{fig:fixedpoints} the four possible fractional cycles on one torus are shown; each has been drawn in only one orbifold sector to gain a better overview. Since the bulk cycles have already been identified and the fractional branes contain half a bulk cycle it is clear how to identify the fractional branes. The green and the cyan ones are those with $|I|=1$. 
From the identification Eq.~(\ref{eq:fracident}) we know that $\eta_1$ and $\eta_4$ must be the branes passing through the origin, so $I=\{1\}$ and $I=\{4\}$ are represented by the green wedge through the fixed point at the origin and 
$I=\{2\}$ and $I=\{3\}$ are represented by the cyan wedge. Two fundamental cycles shifted by both by $0$ or both by $1/2$ recombine to the red brane and two cycles of which one is shifted by $0$ and the other by $1/2$ result in the diagonal through the origin. The only ones left are those with $|I|=3$. These actually describe the same brane as those with $|I|=1$ since by a similarity transformation $E$ and $J$ can be exchanged if this change is compensated by the exchange $u \mapsto -u$ on the moduli space of the bulk cycle.\\
The identification is not yet one-to-one. For each fixed modulus there is the choice of turning on a discrete Wilson line along the brane. When the singularities are "blown-up", this choice corresponds to the orientation of the 2-cycle wrapping the sphere $S^2$. At the level of CFT, this choice is reflected in the charge of a $\mathbb{Z}_2$-symmetry~\cite{DiscreteWilson1,DiscreteWilson2,DiscreteWilson3}. 
\subsection{Orientifold planes}
\subsubsection{Basics}
In the LG action, the reversal of the world-sheet orientation swaps the superspace
coordinates $\theta^{+} \leftrightarrow \theta^{-}$ so that the superpotential term
\begin{eqnarray}
\int d\theta^{+} \theta^{-} W(x),
\end{eqnarray}
picks up a minus sign. This has to be compensated by a holomorphic involution $\sigma$~\cite{orientifold1,orientifold2},
\begin{eqnarray}
W(\sigma x)=-W(x).
\end{eqnarray}
With the boundary fields $\pi_{1,2}, \overline{\pi}_{1,2}$ written out explicitly, the boundary part of the supersymmetry charge reads
\begin{eqnarray}
Q=\left[E^A(x)\pi_1+J^A(x) \overline{\pi}_1\right]_{\sigma=0} - \left[E^B(x)\pi_2+J^B(x) \overline{\pi}_2\right]_{\sigma=\pi}.
\end{eqnarray}
Acting with the parity action on the boundary charge, the two boundary lines get swapped and $x\mapsto \tau x$. It is straightforward to show that this translates to~\cite{orientifold2},
\begin{eqnarray}
Q(x) \mapsto -Q(\sigma x)^T,
\end{eqnarray}
for the factorizations and to,
\begin{eqnarray}
\Phi(x) \mapsto \Phi(\sigma x)^T,
\end{eqnarray}
for the morphisms.
Here, the superscript $(\cdot)^T$ denotes a transposition defined on the $\mathbb{Z}_2$-graded space by,
\begin{eqnarray}
A=\begin{pmatrix} a & b\\c&d\end{pmatrix}
\mapsto A^T=\begin{pmatrix} a^T & -c^T\\b^T&d^T\end{pmatrix}.
\end{eqnarray}
On the subspaces, $(\cdot)^T$ is just the ordinary transpose. Under composition the graded transpose becomes,
\begin{eqnarray}
(AB)^T=(-1)^{|A||B|}B^T A^T.
\end{eqnarray}
When we are looking for invariant branes under the orientifold action $\sigma$ we must of course again allow for similiarity transformations, i.e. the invariance condition translates to,
\begin{eqnarray}
Q(x) = -U(\sigma)Q(\sigma x)^T U(\sigma)^{-1}.\label{eq:orientinv}
\end{eqnarray}
\subsubsection{Geometrical locus}
\begin{figure}[!t]
\begin{center}
\scalebox{0.75}{
\includegraphics{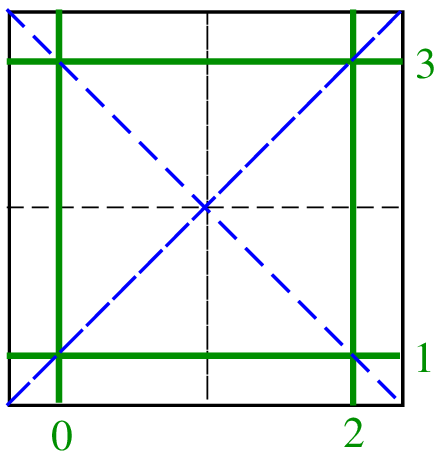}
\includegraphics{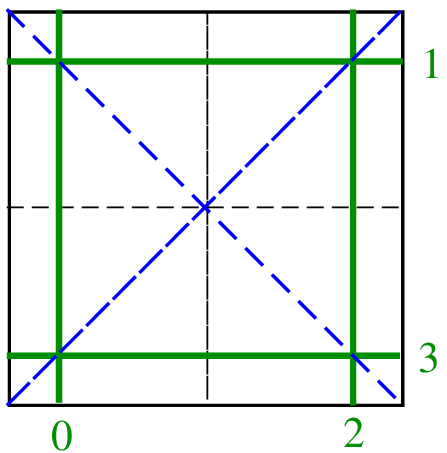}
\includegraphics{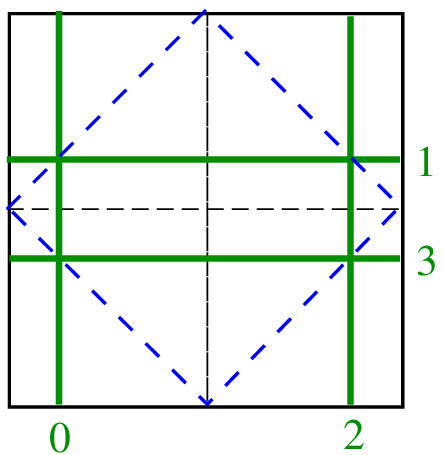}
\includegraphics{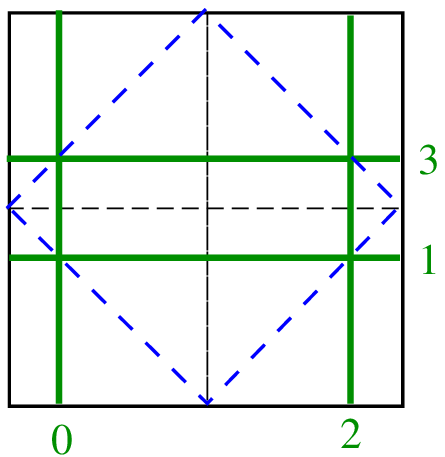}
}
\end{center}
\caption{Branes on the IIA covering space branes under the orbifold actions. The diagrams show the mappings $u\mapsto u$, $u\mapsto -u$, $u\mapsto u+\frac{1}{2}$ and $u\mapsto -u+\frac{1}{2}$.}\label{fig:planes}
\end{figure}

A complete list of orientifold involutions consistent with the complex structure deformation $a_1$ is,
\begin{eqnarray}
\sigma_1^{(n,m)}: (x_1,x_2,z_1) \mapsto (e^{n\pi i(\frac{1}{4}+\frac{n}{2})}x_1,e^{n\pi i(\frac{1}{4}+\frac{m}{2})} x_2,i z_1),\;\;n+m=0\mbox{ mod }2,\\
\sigma_2^{(n,m)}: (x_1,x_2,z_1) \mapsto (e^{n\pi i(\frac{1}{4}+\frac{n}{2})}x_2,e^{n\pi i(\frac{1}{4}+\frac{m}{2})} x_1,i z_1),\;\;n+m=0\mbox{ mod }2,
\end{eqnarray}
with obvious generalization for $x_3,...,x_6,z_2,z_3$. In order
to establish the physical locus of the orientifold planes in the IIA mirror, take a look the orbifold action on the D-branes whose geometrical locus on the $T^2$ is already known.\\
In order to find branes invariant under~(\ref{eq:orientinv}) let us consider the orbifold action on the bulk cycles wrapping parallel to the corrdinate axes. After an appropriate similarity transformation $U$, the factorization is mapped to itself again up to a change in the modulus.
Table~\ref{tab:ori} summarizes the mapping $u\mapsto u'$ under the orientifold action.
\begin{table}[!h]
\begin{center}
\begin{tabular}{|l||c|c|c|c|}
\multicolumn{5}{l}{Action of $\sigma_1^{(n,m)}$:}\\
\hline
$u\mapsto u'(u)$ & $n=0$ & $n=1$ & $n=2$ & $n=3$ \\
\hline
\hline
$m=0$& $u$   &  & $-u+\frac{1}{2}$ &   \\
$m=1$& &  $-u$ &  & $u+\frac{1}{2}$    \\
$m=2$& $-u+\frac{1}{2}$ &   & $u$ &    \\
$m=3$& &  $u+\frac{1}{2}$ &  & $-u$    \\
\hline
\multicolumn{5}{l}{}\\
\multicolumn{5}{l}{Action of $\sigma_2^{(n,m)}$:}\\
\hline
$u\mapsto u'(u)$ & $n=0$ & $n=1$ & $n=2$ & $n=3$ \\
\hline
\hline
$m=0$ &  $-u +\frac{1}{2}+\frac{\tau}{2}$ &  & $u+\frac{\tau}{2}$ &  \\
$m=1$ && $u+\frac{1}{2}+\frac{\tau}{2}$   &  & $-u+\frac{\tau}{2}$ \\
$m=2$ &  $u+\frac{\tau}{2}$               &  & $-u+\frac{1}{2}+\frac{\tau}{2}$ &  \\
$m=3$ && $-u+\frac{\tau}{2}$              &  & $u+\frac{1}{2}+\frac{\tau}{2}$ \\
\hline
\end{tabular}
\caption{Orientifold action on the fundamental bulk cycles.}
\label{tab:ori}
\end{center}
\end{table}
The two involutions $\sigma_1$ and $\sigma_2$ are identical up to a shift in the $\tau$-component. Note that $u$ also contains an implicit $\tau$-component. The complexified modulus $u$ can be decomposed $u=u_{\perp}+\tau u_{\parallel}$. In this decomposition, the real number $u_{\perp}$ describes the location of the brane on the covering space. The real number $u_{\parallel}$ corresponds to a Wilson line. It is known that the orientifold map can induce a minus sign. In addition, there is the optional shift by $1/2$ (modulo 1) in the Wilson line. In principle it is also known that a $\mathbbm{Z}_2$ action on a periodic coordinate could be added but I am not aware of any example with branes in the bulk. For fractional branes see e.g.~\cite{DiscreteWilsonpractical}. For the fractional branes here, this shift corresponds to turning the discrete Wilson line on or off. Geometrically speaking this corresponds to the orientation of the cycles wrapping the "blowed-up" singularities.\\\\
Table~\ref{tab:ori} is not yet the full story, however. It must be taken into account that the action of the similarity transformation $U$ acts on the orbifold generator as well by,
\begin{eqnarray}
U^{-1}\gamma^i U=\gamma^{i+1}.
\end{eqnarray}
That means the orientifold action changes the orbit by 1. For the diagonal bulk cycles the orbit is left invariant. One needs four of the fundamental bulk cycles at generic points to get an invariant action, at special points two suffice. For the diagonal cycles at most two are needed, some are invariant by themselves. The data of the bulk cycle in the table together with the last equation is displayed graphically in Fig.~\ref{fig:planes}. The first two graphs differ just by the orbits of the branes and so do the two others. The difference is perhaps just the orientation of the instanton bounded by the branes.
\section{Stability of brane configurations}
Matrix factorizations are endowed with a grading. From the R-charges of the matrix entries one computes a (diagonal) R-matrix. In terms of this matrix the orbifold generator $\gamma_Q$ reads,
\begin{eqnarray}
\gamma_Q=\sigma e^{\pi i R} e^{-i \pi \lambda^Q_k}.
\end{eqnarray}
The phases $\lambda^Q_k$ of the factorization $Q$, which are restricted by $\gamma_Q^4=1$ in our $\mathbbm{Z}_4$-symmetric case, give rise to the four different orbits of the branes.
These orbifold copies are pairwise identical for the even-dimensional orbifold group, so effectively only two copies remain.
Up to here, the phase is determined only modulo 2. While this is sufficient when dealing with just an isolated D-brane, the ambiguity must be taken into account when analyzing an entire brane configuration including the open strings stretching between the branes~\cite{doug1,doug2}.
It is necessary to lift the phase to a real number and associate with every brane $Q$ a grading $n$ which is the integer offset of the lifted phase. This refinement induces an analogous grading $m$ on each morphism $\Phi_{(P,Q)}$,
\begin{eqnarray}
m=\lambda^Q-\lambda^P+q_{\Phi}.\label{eq:morphgrading}
\end{eqnarray}
where $q_\Phi$ is the string's R-charge. The states are bosonic for odd $m$ and fermionic for even $m$. It has been argued that the difference in the grades $\Delta\lambda=\lambda^Q-\lambda^P$ measures the squared mass of the fermionic state $\psi_{(P,Q)}$ stretching between the two branes~\cite{doug3,aspin1,aspin2,jockers}. For $\Delta\lambda>0$ the fermion is massive, for $\Delta\lambda<0$ it becomes tachyonic.
In order to obtain a massless fermionic spectrum, the grades of all factorizations must therefore be identical. Eq.~(\ref{eq:morphgrading}) constrains the fermionic states to have even integer R-charge $q_{\Phi}$. Since each $T^2$ contributes 1 to the total central charge $\hat c$ and the morphisms have to satisfy the unitarity bound $0 \le q_{\Phi} \le \hat c$, the open string charge must be zero for all strings in a $T^2$ construction and $0$ or $2$ for all states in $T^4$ and $T^6$ constructions. The absence of tachyons indicates the absence of manifolds with lower volume and identical combined wrapping number, so the configuration should be stable.\\
\begin{figure}
\begin{center}
\includegraphics{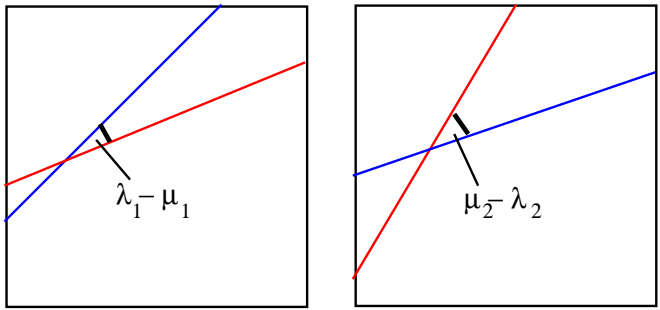}
\end{center}
\caption{The $T^2 \times T^2$ covering space with two factorizable 2-cycles.}\label{fig:stab}
\end{figure}
To gain some confidence in the findings, let us briefly compare them to results for factorizable cycles obtained by conventional methods. A 1-cycle wrapping $T^2$ encloses an angle with the $x$-axis, which corresponds to $\lambda$ mod $1$ in units of $\pi$. The angle between two 1-cycles
corresponds to the R-charge of the open string stretching between the two branes. Above it was argued that this charge must vanish, so the only stable configuration consists of parallel branes. This is also intuitively clear since in the 2D case one can easily visualize the dynamical recombination process into volume minimizing branes. The conclusion in is complete agreement with conventional intersecting brane modeling~\cite{IBMstab} where the light scalar field has a mass squared of,
\begin{eqnarray}
m^2 \alpha' = -\frac{1}{2 \pi}\theta,
\end{eqnarray}
which is tachyonic for every non-zero angle $\theta$. For factorizable cycles on the $T^4$, there are two scalars located at the two intersection angles $\theta_1$ and $\theta_2$ on the two tori, with masses,
\begin{eqnarray}
m_1^2 \alpha' = -\frac{1}{2 \pi}(\theta_1-\theta_2) \qquad m_2^2 \alpha' = -\frac{1}{2 \pi}(\theta_2-\theta_1).
\end{eqnarray}
There is a "line of stability" when the angles are identical $\theta_1=\theta_2$. The same results are obtained from the above category theory considerations: Tensor together two 1-cycles with phases $\lambda_{1,2}$ and obtain a factorizable 2-cycle on $T^4$ with $\lambda=\lambda_1+\lambda_2$, then construct another 2-cycle with phase $\mu=\mu_1+\mu_2$ in the same manner. The satability condition $\lambda=\mu$ implies,
\begin{eqnarray}
(\lambda_1-\mu_1)-(\mu_2-\lambda_2)=0.
\end{eqnarray}
The difference between the phases are just the angles (or the negatives of it) on the tori as is illustrated in Fig.~\ref{fig:stab}. Finally, for the $T^6$ four potential tachyons are obtained in the intersecting brane literature. The moduli space of the angles is a tetrahedron enclosing all stable configurations. Category theory gives,
\begin{eqnarray}
\lambda_1+\lambda_2+\lambda_3=\mu_1+\mu_2+\mu_3.
\end{eqnarray}
The r.h.s. can again be brought to the left in order to group the angle differences together. It then reads,
\begin{eqnarray}
\pm \theta_1 \pm \theta_2\pm \theta_3=0.
\end{eqnarray}
The signs of $\theta_i$ depend on whether $\lambda_i > \mu_i$ or $\lambda_i < \mu_i$. The equation defines one face of the tetrahedron. The reason why only one side of the entire tetrahedron is obtained becomes clear when one remembers that by construction the LG theory preserves $\mathcal{N}=1$ supersymmetry for a single brane, whereas the angle configurations within the tetrahedron are not supersymmetric. On its sides $\mathcal{N}=1$ SUSY is preserved and its edges preserve even $\mathcal{N}=2$ (which is the case when one $\theta_i$ vanishes). The choice of phase in the LG theory therefore amounts to the choice of which supersymmetry is preserved. The extension to Q-SUSY, that is models which preserve a different unbroken $\mathcal{N}=1$ SUSY at each brane intersection, follows an analogous line of argument.\\
To sum up, in the framework of matrix factorizations one can start with a brane configuration without global supersymmetry (that is, each brane by itself preserves a different $\mathcal{N}=1$) and this configuration then recombines dynamically to a stable ground state with a common supersymmetry. Stability of a massless tachyon-free spectrum is indicated by grades $\lambda$ which are identical for all branes. The grading is easy to compute in the matrix factorization framework in contrast to the difficulty of determining stability for non-factorizable cycles by conventional IBM methods. This is one of the reasons why typically only factorizable branes are studied although it is well-known that non-factorizable branes are generically unavoidable after brane recombinations.
\section{The $T^6 / \mathbbm{Z}_2 \times \mathbbm{Z}_2$-orbifold}
In the intersecting brane literature the $T^6 / \mathbbm{Z}_2 \times \mathbbm{Z}_2$-orientifolds are the phenomenologically most important candidates, therefore it would be desirable to be able to construct these models with matrix factorizations as well.\\
The number of three-cycles which are not projected out depending on the orbifold group $\Gamma$ are:
\begin{center}
\begin{tabular}{|c||c|c|c|}
\hline
$\Gamma$ & $\mathbbm{Z}_4\times \mathbbm{Z}_4$ & $\mathbbm{Z}_2\times \mathbbm{Z}_2$\\
\hline
$b_3$ & 2 & 8\\
\hline
\end{tabular}
\end{center}
The two remaining 3-cycles $\Pi_{1,2}$ for the first case were identified with the two orbifold copies of the matrix factorizations,
\begin{eqnarray}
\Pi_i \simeq (Q^{(3)} \otimes Q^{(3)} \otimes Q^{(3)})_i, \qquad i=1\mbox{ or }2.
\end{eqnarray}
The notation means that the three factorizations are first tensored and then given an orbifold label 1 or 2. Since the orbifold generators for $\Gamma=\mathbbm{Z}_2\times \mathbbm{Z}_2$ are precisely twice those of $\Gamma=\mathbbm{Z}_4\times \mathbbm{Z}_4$, it suffices to orbifold by the squared generator to obtain the desired group action. For completeness here the eight remaining 3-cycles:
\begin{eqnarray}
\pi_{i,j+2,k+4} \simeq Q^{(3)}_i \otimes Q^{(3)}_j \otimes Q^{(3)}_k, \qquad i,j,k=1\mbox{ or }2.
\end{eqnarray}
In other words, the factorization can be thought of as possessing one orbifold label for each tensored element, where the label denotes the fundamental cycle on the respective torus.\\\\
From here one could continued by combining these branes through tachyon condensation and derive their low-energy effective potential as demonstrated in~\cite{knappomer1}. From their R-charges one can follow~\cite{modulistab} and check, if the anomaly cancellation condition is satisfied.
\section{Acknowledgements}
This work could not have been pursued without the kind support of Cumrun Vafa, Johannes Walcher and especially Wolfgang Lerche.
Furthermore, I appreciate Fernando Marchesano's patient explanations on intersecting brane modeling. 
\appendix
\section{Theta Functions}
The convention used for the Jacobi Theta functions are as follows:
\begin{equation}
\Theta\left[\begin{array}{c}c_1\\c_2\end{array}\right](u,\tau)=\sum_{m\in\mathbb{Z}}q^{(m+c_1)^2/2}e^{2\pi i(u+c_2)(m+c_1)}
\qquad q=e^{2\pi i \tau}.
\end{equation}
\begin{eqnarray}
\Theta_1(u,\tau)\equiv\Theta\left[\begin{array}{c}\frac{1}{2}\\\frac{1}{2}\end{array}\right](u,\tau)\quad \Theta_2(u,\tau)\equiv\Theta\left[\begin{array}{c}\frac{1}{2}\\0\end{array}\right](u,\tau)\\
\Theta_3(u,\tau)\equiv\Theta\left[\begin{array}{c}0\\0\end{array}\right](u,\tau)\quad\Theta_4(u,\tau)\equiv\Theta\left[\begin{array}{c}0\\\frac{1}{2}\end{array}\right](u,\tau)
\end{eqnarray}

\end{document}